# Approximate virtual quantum broadcasting

Matthew Simon Tan,[1, *] Davit Aghamalyan,[2, †] and Varun Narasimhachar[3, ‡]

[1]*National University of Singapore (NUS), 21 Lower Kent Ridge Rd, University Hall, Republic of Singapore 119077*
[2]*Singapore University of Technology and Design (SUTD), 8 Somapah Rd, Republic of Singapore 487372*
[3]*Institute of High Performance Computing, Agency for Science,
Technology and Research (A\*STAR), 1 Fusionopolis Way, Republic of Singapore 138632*
(Dated: March 19, 2026)

The no-broadcasting theorem, a fundamental limitation on the communication of quantum information, holds that a physical process cannot broadcast copies of an unknown quantum state to two or more receivers. Recent work has explored ways of circumventing this limitation using "virtual" implementations of non-physical processes using measurement and data-processing on statistical samples of the unknown input. However, the statistical fluctuations of this data degrades the virtual copies so much that the protocol effectively depletes, rather than proliferate, the sample size—thereby rendering it worse than the "naive" approach of splitting the given sample and sending a subsample to each receiver. In this work, we circumvent this flaw by allowing a small amount of systematic bias in the broadcast data, resulting in approximate virtual copies. We provide efficient semidefinite programs (SDP's) to determine the minimum sample size required to keep the approximation error below a desired threshold and vice versa. For reasonably small error values, we find approximate virtual broadcasting to be viable with sample sizes smaller than naive sample-splitting would demand. Along the way, we prove several symmetry-based simplifications to the problem, allowing optimal approximate broadcasting to be characterized in terms of the simple class of depolarizing channels.

## I. INTRODUCTION

Quantum information processing promises computational [1, 2] and communication [3, 4] capabilities that are classically unattainable, yet it is constrained by a family of fundamental no-go theorems. Iconic among these are the results of no-cloning [5, 6] and no-broadcasting [7, 8], where the latter prohibits any deterministic physical process (so-called *quantum channel*) from distributing perfect copies of an unknown quantum state to two or more receivers (or clients). These constraints extend beyond the linearity of quantum mechanics and establish fundamental limits on physically realizable information-processing protocols.

This limitation can be circumvented in different ways, many of which essentially amount to compromising on determinism (i.e. a perfect success rate). One such recourse is so-called *virtual implementation*, viable for certain processes. In technical terms, these are the processes modeled by Hermitian-preserving, trace-preserving (HPTP) linear maps, even though they may fail to be channels, which are modeled by completely-positive trace-preserving (CPTP) maps. Such processes can be simulated via Monte Carlo sampling of channels, followed by post-processing of measurement data [9, 10]. Virtual implementations have been found for a variety of tasks, including resource distillation [11–13], state cloning [14], channel simulation [15, 16], and more recently, broadcasting [17, 18].

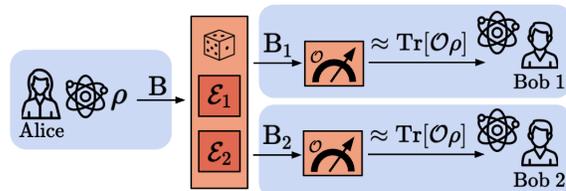

FIG. 1. **Summary of approximate virtual quantum broadcasting.** Copies of a state $\rho$ is sent through the channels $\mathcal{E}_1$ and $\mathcal{E}_2$ and then measured with the observable $\mathcal{O}$. The results are then classically post-processed to obtain an estimation of $\text{Tr}[\mathcal{O}\rho]$. Allowing some bias allows us to use less copies of $\rho$ than the naive strategy of preparation and distribution.

However, virtual implementation necessarily acts on a statistical sample consisting of many copies of an input state, and produces measurement data corresponding to many copies of the desired output. The performance of the implementation is measured by the precision to which the data estimates the true values of observables, and the size of the sample required to attain a desired precision level (also called *sample complexity*). For the case of the broadcasting task, where the virtual implementation is of a process that itself has the desired effect of proliferating the sample, this raises a crucial question: does the sample complexity overhead of virtual implementation still leave room for the broadcast to be a net proliferation? The answer turns out to be no [17, 19–21][1], calling into question the practical utility of virtual broadcasting.

---

* e0726673@u.nus.edu
† davagham@gmail.com
‡ varun.achar@gmail.com

[1] The operational conditions imposed on the candidate broadcast-

In this work, we show that the sample complexity can be brought down to practical levels if we allow for *approximate* virtual implementation. Our main idea is to allow for a small systematic error in the output received by each client, yielding virtual copies of a state slightly different from the given input. We formulate an efficiently computable semidefinite program (SDP) [22–24] for minimizing the sample complexity for a given approximation threshold, as well as one for the inverse problem of minimizing the approximation error for a given sample complexity budget. Using these, we show that allowing an error as small as 15% (compared to random noise) enables virtually broadcasting qubits (2-dimensional quantum states) at a rate more efficient than the naive sample-splitting approach. Using an analytical upper bound on the minimum error in the inverse problem, we also show that states of arbitrary dimension can be broadcast sample-efficiently by incurring errors smaller than 42%.

In the course of our technical derivations, we also find some symmetries in the approximate virtual broadcasting problem, which serve to both simplify the mathematical treatment and afford insight into the problem's structure. In particular, we demonstrate that no optimality is lost by restricting to maps satisfying unitary covariance; this, in turn, leads to the operational simplification that optimal approximate broadcasting is possible with the effective virtual channel from the source to each client a *depolarizing channel*. This is an extensively studied and practically important class of channels, hinting at the practical viability of virtual approximate broadcasting.

This paper is organized as follows. In Section III, we outline the framework of virtual broadcasting and formalize the sample complexity overhead. In Section IV, we introduce the framework of approximate virtual broadcasting, and formulate our main problem of minimizing the sample complexity overhead for keeping the error below a given threshold; we then solve the problem through a series of symmetry-based simplifications. In Section V, we consider the inverse problem of minimizing the error for a given sample complexity budget, resulting in explicit upper bounds on the error incurred for sample-efficient virtual broadcasting. Finally, Section VI is devoted to conclusions and future directions.

## II. NOTATION

In this work, we restrict to the case of broadcasting from one source to two clients. We will call the source $B$ and the clients $B_1$ and $B_2$. Each of these systems is a qudit, i.e., associated with a Hilbert space $\mathcal{H}$ of finite dimension $d$. We denote by $\mathcal{B}(\mathcal{H})$ the space of bounded

---

ing protocols vary in these works, but the negative answer holds even under the most relaxed conditions, considered in Ref. [17].

---

linear operators acting on $\mathcal{H}$. Quantum states of a qubit are associated with density operators $\rho \in \mathcal{B}(\mathcal{H})$, satisfying $\rho \geq 0$ and $\text{Tr}[\rho] = 1$; denote the space of all such density operators by $\mathcal{D}(\mathcal{H})$. Qudit-to-qudit quantum channels are associated with linear endomorphisms of $\mathcal{B}(\mathcal{H})$ that are completely positive (CP) and trace-preserving (TP). We fix a canonical orthonormal basis $\{|i\rangle\}$ for $\mathcal{H}$, and define $\Gamma := \sum_i |ii\rangle \in \mathcal{H} \otimes \mathcal{H}$. For a channel $\mathcal{E}$, its Choi operator (relative to this basis) is $\mathcal{J}^{\mathcal{E}} := (I \otimes \mathcal{E})\Gamma$. Finally, we will omit qudit labels on operator and superoperator symbols where the associated qudits are clear from the context. When we do use explicit labels, we adopt the following convention: if $L$ is a one-qudit linear operator, then $L_B$ will denote $L$ instantiated on the space of system $B$; similarly, for a two-qudit operator, $M_{BB_1}$ will denote $M$ instantiated on the spaces of $BB_1$, etc. In such cases, we will not use the convention where $M_B$ denotes the corresponding marginal, instead denoting the latter explicitly as $\text{Tr}_{B_1} M_{BB_1}$, or simply $\text{Tr}_{B_1} M$.

## III. QUANTUM BROADCASTING

In this section, based on Ref. [20], we review the framework of virtual quantum broadcasting, the sample complexity of virtual implementation, and the SDP for computing the latter. We also formalize the naive sample-splitting strategy, against which virtual broadcasting will be compared.

**Definition 1** (Broadcasting map). Let $\mathcal{E}_{B \to B_1 B_2}$ be a Hermitian-preserving, trace-preserving (HPTP) map. We call it a broadcasting map if it satisfies

$$\text{Tr}_{B_1}[\mathcal{E}(\rho_B)] = \rho_{B_2} \text{ and} \tag{1}$$

$$\text{Tr}_{B_2}[\mathcal{E}(\rho_B)] = \rho_{B_1} \tag{2}$$

for all $\rho \in \mathcal{D}(\mathcal{H})$.

Phrasing this condition in the language of the Choi–Jamiołkowski isomorphism [25, 26], we obtain the following.

**Proposition 1** ([17]). *An HPTP map $\mathcal{E}$ is a broadcasting map if and only if its Choi operator $\mathcal{J}^{\mathcal{E}}$ satisfies the following conditions:*

$$\text{Tr}_{B_2}[\mathcal{J}^{\mathcal{E}}] = \Gamma_{BB1} \text{ and} \tag{3}$$

$$\text{Tr}_{B_1}[\mathcal{J}^{\mathcal{E}}] = \Gamma_{BB2}. \tag{4}$$

Recall that HPTP maps are associated with maps that are not physical quantum channels, but which can be virtually implemented through sampling quantum channels, measuring, and post-processing. However, this simulation process induces an increase in the number of samples needed to estimate observables to a desired precision. We shall now define and examine the sample complexity overhead $S(\mathcal{E})$ implementing an HPTP map $\mathcal{E}$ virtually.

Suppose we would like to estimate with probability $1 - \delta$ the value of $\text{Tr}[\mathcal{O}\mathcal{E}(\rho)]$ up to an error $\varepsilon$. By the Hoeffding inequality [27], we would require at least $\frac{M^2}{\varepsilon^2}\ln\left(\frac{2}{\delta}\right)$ copies of $\mathcal{E}(\rho)$, where $M$ is the difference between the largest and smallest eigenvalues of $\mathcal{O}$. Given an HPTP channel $\mathcal{E}$, we associate with it a decomposition $(x\mathcal{E}_+, y\mathcal{E}_-)$, such that $\mathcal{E} = x\mathcal{E}_+ - y\mathcal{E}_-$, which we call a virtual protocol. This protocol can be implemented in the following manner. We have by linearity of trace that

$$\text{Tr}[\mathcal{O}\mathcal{E}(\rho)] = (x+y)\left(p_+ \text{Tr}[\mathcal{O}\mathcal{E}_+(\rho)] - p_- \text{Tr}[\mathcal{O}\mathcal{E}_-(\rho)]\right), \quad (5)$$

where $p_+ := \frac{x}{x+y} =: 1 - p_-$. By sampling $\mathcal{E}_\pm$ with respective probability $p_\pm$ and then scaling the resulting values by a factor of $(x+y)$, we can obtain estimates on $Tr[\mathcal{O}\mathcal{E}(\rho)]$. However, we are now effectively estimating the scaled observable $(x+y)\mathcal{O}$, hence the range of eigenvalues is also scaled by a factor of $(x+y)$. Therefore, Hoeffding's inequality would now entail $\frac{M^2(x+y)^2}{\varepsilon^2}\ln\left(\frac{2}{\delta}\right)$ samples. Thus, we can see that the number of samples required to estimate an observable is determined by the choice of decomposition of the virtual channel $\mathcal{E}$ that we make. This leads to the following definition for the minimum overhead:

**Definition 2** (Sample complexity overhead). For an HPTP map $\mathcal{E}$, we define the sample complexity overhead via the following optimization problem.

$$\begin{aligned}
\mathcal{S}(\mathcal{E}) := \min \quad & (x+y)^2 \\
\text{s.t.} \quad & \mathcal{E}_1, \mathcal{E}_2 \in \text{CPTP}, \quad \mathcal{E} \in \text{HPTP}; \\
& x, y \geq 0; \\
& x - y = 1; \\
& \mathcal{E} = x\mathcal{E}_1 - y\mathcal{E}_2;
\end{aligned} \quad (6)$$

Since minimizing $(x+y)^2$ amounts to minimizing $(x+y)$, this can be framed as the following SDP using Choi operators:

$$\begin{aligned}
\min \quad & x+y \\
\text{s.t.} \quad & \mathcal{J}^\mathcal{E} = \mathcal{J}^{\mathcal{E}_1} - \mathcal{J}^{\mathcal{E}_2}, \\
& \text{Tr}_{B_1 B_2}[\mathcal{J}^{\mathcal{E}_1}] = xI_B, \\
& \text{Tr}_{B_1 B_2}[\mathcal{J}^{\mathcal{E}_2}] = yI_B, \\
& \mathcal{J}^{\mathcal{E}_1} \geq 0, \quad \mathcal{J}^{\mathcal{E}_2} \geq 0. \\
& x - y = 1.
\end{aligned} \quad (7)$$

In the case of 1-to-2 broadcasting, since $\mathcal{O}_1 = \mathcal{O}_2$, in order to beat the naive strategy, we must have that the sample overhead $n(x+y)^2 < 2n$. In other words, the above SDP must have an optimal value strictly less than $\sqrt{2}$. Since the optimal value turns out to be $\frac{3d-1}{d+1}$ [17], we know that this is not possible for any $d \geq 2$. Consequently, there does not exist a 1-to-2 broadcasting map that is able to (virtually) outperform the naive strategy. This motivates us to explore approximate virtual broadcasting, which we will now formalize.

## IV. APPROXIMATE VIRTUAL BROADCASTING

In this section, we define the task of approximate virtual broadcasting and express its associated sample complexity as an SDP. We then show that the optimal solution can be assumed to be unitarily covariant without loss of optimality, drastically simplifying the SDP. This simplified form not only makes the sample complexity amenable to numerical analysis but is also of independent interest, as it allows us to obtain an elegant characterization of optimal strategies for this task.

**Definition 3** (Approximate broadcasting map). Let $\mathcal{E}_{B \to B_1 B_2}$ be an HPTP map, and $(a, b) \in [0, 1]^2$. Then $\mathcal{E}$ is an $(a, b)$-approximate broadcasting map if $\mathcal{E}_{B \to B_1 B_2}$ satisfies

$$\begin{aligned}
\frac{1}{2} \left\| \text{Tr}_{B_2} \circ \mathcal{E} - id_{B \to B_1} \right\|_\diamond &\leq a \text{ and} \\
\frac{1}{2} \left\| \text{Tr}_{B_1} \circ \mathcal{E} - id_{B \to B_2} \right\|_\diamond &\leq b.
\end{aligned} \quad (8)$$

In the case that $a = b = \delta$, we call it a $\delta$-approximate broadcasting map as a shorthand. We denote the set of $(a, b)$-approximate broadcasting maps by $\mathcal{B}_{(a,b)}$.

Similar to [15], we choose the diamond norm as a figure of merit on each marginal as it because of its interpretation as a measure of distinguishability between quantum channels and because it measures the worst-case error in estimating $\text{Tr}[\mathcal{O}\rho]$. This can be seen in the following way:

Suppose we have two quantum states $\rho$ and $\sigma$, then $\frac{1}{2} \|\rho - \sigma\|_1$ is the maximum success probability in distinguishes the two states $\rho$ and $\sigma$ in a single shot measurement scenario. Therefore, the metric induced by the diamond norm can be interpreted as the maximum success probability that distinguishes two quantum channels. Furthermore, consider a $\delta$-approximate broadcasting map $\mathcal{E}$ and one if its marginal channels $\text{Tr}_{B_i} \circ \mathcal{E}$, such that $\| \text{Tr}_{B_i} \circ \mathcal{E} - \text{id}_{B \to B_i} \|_\diamond \leq \delta$ Then we have, for any density operator $\rho$:

$$\begin{aligned}
&|\text{Tr}[\mathcal{O}\rho] - \text{Tr}[\mathcal{O} \text{Tr}_{B_i}\{\mathcal{E}(\rho)\}]| \\
&= |\text{Tr}[\mathcal{O}(\rho - \text{Tr}_{B_i}\{\mathcal{E}(\rho)\})]| \\
&\leq \|O\|_\infty \|\rho - \text{Tr}_{B_i}\{\mathcal{E}(\rho)\}\|_1 \\
&\leq \|\mathcal{O}\|_\infty \|\text{id}_{B \to B_i} - \text{Tr}_{B_i} \circ \mathcal{E}\|_\diamond
\end{aligned} \quad (9)$$

where the first inequality comes from Hölder's and the second inequality comes from the definition of the diamond norm. So for a fixed observable $\mathcal{O}$, the diamond norm gives us control on the worst-case error in estimating $\text{Tr}[\mathcal{O}\rho]$.

To quantify the resources needed to implement the approximate virtual broadcasting map, we introduce the first of two main quantities that we will study. The first quantity determines the sample complexity associated with a particular approximate broadcasting task. More specifically, given two error thresholds $(a, b)$ to be

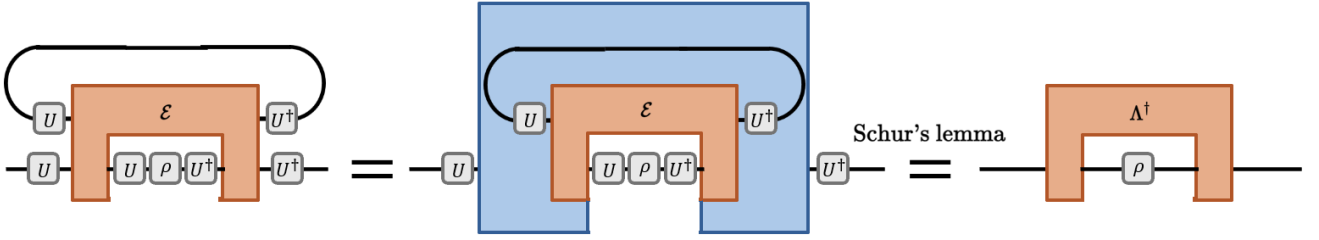

FIG. 2. **Visual summary of the proof of Theorem 1, using tensor diagrams.** We first twirl the broadcasting map $\mathcal{E}$ over the unitary group. Focusing on one marginal channel, we see that this makes the channel unitary covariant. Thus we can invoke Schur's lemma, and reduce the channel to one of the parameterized depolarizing channels $\Lambda^t$.

achieved on the marginals, we can look over all possible $(a,b)$-approximate broadcasting maps, and select for the map $\mathcal{E}$ that achieves the lowest sample complexity $S(\mathcal{E})$. That is we give the following definition

**Definition 4.** Let $(a,b) \in [0,1]^2$ be a pair of error thresholds. We define the sample complexity associated with the task of $(a,b)$-broadcasting as follows:

$$S(a,b) := \min \quad S(\mathcal{E}) \\ \text{s.t.} \quad \mathcal{E} \in \mathcal{B}_{(a,b)} \tag{10}$$

Recall from the above discussion that if the sample complexity associated with implementing an HPTP map is greater than 2, then a naive prepare-and-distribute approach is more efficient in the number of samples of the state $\rho$ required for a similar level of accuracy. Therefore, we will say that an approximate broadcasting map $E$ satisfies sample efficiency (SE) if $S(\mathcal{E}) < 2$.

The primary question we are interested in is if there exists an $(a,b)$-approximate broadcasting maps satisfying SE, and furthermore, how we can characterize the landscape of tradeoffs between the error thresholds and the sample complexity of its associated broadcasting maps.

For any $\delta \in [0,1]$, we abbreviate $S(\delta, \delta)$ as $S_\delta$. Note that the optimization problem associated with $S(a,b)$ can be computed via SDP. In particular, $\sqrt{S(a,b)}$ is an SDP, and since the obejctive is non-negative throughout the feasible set, we can compute first $\sqrt{S(a,b)}$, then square the outcome. Since the square rooted quantity has a linear objective function, it is often more expedient to work with it; in the remainder, we denote $\tilde{S}(a,b) := \sqrt{S(a,b)}$. To see that $\tilde{S}(a,b)$ can be computed via SDP we can again use the Choi–Jamiołkowski isomorphism, to obtain

$$\tilde{S}(a,b) = \min \quad x + y \\ \text{s.t.} \quad \frac{1}{2} \|\text{Tr}_{B_2} \circ \mathcal{E} - \text{id}_{B \to B_1}\|_\diamond \leq a, \\ \frac{1}{2} \|\text{Tr}_{B_1} \circ \mathcal{E} - \text{id}_{B \to B_2}\|_\diamond \leq b, \\ \mathcal{E} = \mathcal{E}_1 - \mathcal{E}_2, \\ \text{Tr}_{B_1 B_2}[\mathcal{J}^{\mathcal{E}_1}] = x I_B, \\ \text{Tr}_{B_1 B_2}[\mathcal{J}^{\mathcal{E}_2}] = y I_B, \\ x - y = 1 \\ \mathcal{J}^{\mathcal{E}_1} \geq 0, \quad \mathcal{J}^{\mathcal{E}_2} \geq 0. \tag{11}$$

The above is a manifestly an SDP. Note that a feasible point is given by the tuple $(\mathcal{J}^{\mathcal{E}_1}, \mathcal{J}^{\mathcal{E}_2}, x, y)$, but since $x$ and $y$ are uniquely determined by the channels $\mathcal{E}_1$ and $\mathcal{E}_2$, we can for brevity refer to feasible points as pairs $(\mathcal{J}^{\mathcal{E}_1}, \mathcal{J}^{\mathcal{E}_2})$.

We first focus on simplifying the optimization problem associated with $\tilde{S}(a,b)$. Consider a fixed total "budget" for the error, as in, we consider all $(a,b)$-approximate broadcasting tasks such that $a + b = c$ for some fixed $c$. Through numerics, we can observe that the set of error thresholds on each curve $c = a + b$ that minimize the sample complexity are given by the diagonal $(\frac{c}{2}, \frac{c}{2})$. This suggests that if we aim to minimize the worst-case error on each marginal, it suffices to look only at the diagonals. This intuition is formalized by the following proposition.

**Proposition 2.** *Let $\tilde{S} : [0,1]^2 \to \mathbb{R}_{\geq 0}$ be the function taking each pair of error thresholds $(a,b)$ to the sample complexity associated with $(a,b)$-approximate broadcasting. Then $\tilde{S}$ is jointly convex, and $\tilde{S}(a,b) = \tilde{S}(b,a)$.*

*Proof.* We first show that the function is symmetric in its inputs. Take any pair $(\mathcal{J}^{\mathcal{E}_1}, \mathcal{J}^{\mathcal{E}_2})$ feasible for the $\tilde{S}(a,b)$. Then define $\mathcal{F}_1 := \pi \circ \mathcal{E}_1$ where $\pi$ is the unique linear extension of the bipartite channel given by $\pi(\rho \otimes \sigma) = \sigma \otimes \rho$ for all $\rho, \sigma \in \mathcal{D}(\mathcal{H})$. Similarly define $\mathcal{F}_2 = \pi \circ \mathcal{E}_2$. Then $(\mathcal{J}^{\mathcal{F}_1}, \mathcal{J}^{\mathcal{F}_2})$ is feasible for the optimization problem associated with $\tilde{S}(b,a)$, with the same objective value. Similar feasible points for the optimization problem associated with $\tilde{S}(b,a)$ have corresponding points for the optimization problem associated with $\tilde{S}(a,b)$ with the same objective value. Thus the function is symmetric.

It remains to show joint convexity. We need to show that for any two $(a,b),(a',b') \in [0,1] \times [0,1]$

$$\lambda \tilde{S}(a,b) + (1-\lambda)\tilde{S}(a',b')
\geq \tilde{S}(\lambda a + (1-\lambda)a', \lambda b + (1-\lambda)b') \text{ for } \lambda \in [0,1] \quad (12)$$

Suppose $(\mathcal{J}^{\mathcal{E}_1}, \mathcal{J}^{\mathcal{E}_2})$ optimizes the sample overhead for $(a,b)$, and $(\mathcal{J}^{\mathcal{E}'_1}, \mathcal{J}^{\mathcal{E}'_2})$ optimizes the sample overhead for $(a',b')$. Then, we claim that the point $(\lambda \mathcal{J}^{\mathcal{E}_1} + (1-\lambda)\mathcal{J}^{\mathcal{E}'_1}, \lambda \mathcal{J}^{\mathcal{E}_2} + (1-\lambda)\mathcal{J}^{\mathcal{E}'_2})$ is feasible for the error thresholds $(\lambda a + (1-\lambda)a', \lambda b + (1-\lambda)b')$. Indeed, one can check if $\mathcal{J}^{\mathcal{E}} = \mathcal{J}^{\mathcal{E}_1} - \mathcal{J}^{\mathcal{E}_2}$ and $\mathcal{J}^{\mathcal{E}'} = \mathcal{J}^{\mathcal{E}'_1} - \mathcal{J}^{\mathcal{E}'_2}$, indeed, we have

$$\begin{aligned}
&\|id - \text{Tr}_{B_1}(\lambda \mathcal{E} + (1-\lambda)\mathcal{E}')\|_\diamond \\
&= \|\lambda(id - \text{Tr}_{B_1} \circ \mathcal{E}) + (1-\lambda)(id - \text{Tr}_{B_1} \circ \mathcal{E}')\|_\diamond \\
&\leq \lambda \|id - \text{Tr}_{B_1} \circ \mathcal{E}\|_\diamond + (1-\lambda) \|id - \text{Tr}_{B_1} \circ \mathcal{E}'\|_\diamond \\
&\leq \lambda a + (1-\lambda)a'
\end{aligned} \quad (13)$$

and similarly for the other marginal. Finally, $\text{Tr}_{B_1 B_2}[\lambda \mathcal{J}^{\mathcal{E}_1} + (1-\lambda)\mathcal{J}^{\mathcal{E}'_1}] = \{\lambda x + (1-\lambda)x'\}I_B$, and similarly, $\text{Tr}_{B_1 B_2}[\lambda \mathcal{J}^{\mathcal{E}_2} + (1-\lambda)\mathcal{J}^{\mathcal{E}'_2}] = \{\lambda y + (1-\lambda)y'\}I_B$. Thus the point

$$(\lambda \mathcal{J}^{\mathcal{E}_1} + (1-\lambda)\mathcal{J}^{\mathcal{E}'_1}, \lambda \mathcal{J}^{\mathcal{E}_2} + (1-\lambda)\mathcal{J}^{\mathcal{E}'_2})$$

is feasible for the SDP, and we have

$$\begin{aligned}
&\tilde{S}(\lambda a + (1-\lambda)a', \lambda b + (1-\lambda)b') \\
&\leq [\lambda x + (1-\lambda)x'] + [\lambda y + (1-\lambda)y'] \\
&= \lambda(x+y) + (1-\lambda)(x'+y') \\
&= \lambda \tilde{S}(a,b) + (1-\lambda)\tilde{S}(a',b')
\end{aligned} \quad (14)$$

Therefore, we have that the map $\tilde{S}(a,b)$ is both jointly convex and symmetric. $\square$

With this proposition, we see that along a fixed constraint curve the minimum sample complexity is attained at the balanced point. Since our ultimate aim is to minimizes the worst-case error on each marginal, we can restrict the discussion to considering only balanced error thresholds. Thus, for the remainder of this section we focus our attention on $\tilde{S}_\delta$.

We now show through a unitary twirling argument that, without loss of optimality, we can restrict the feasible set of this $\tilde{S}(\delta)$ to the set of unitary covariant channels. We briefly recall the definition of unitary twirling for completeness.

**Definition 5.** Let $U(d)$ be the $d$-dimensional unitary group and let $\mu_H$ denote the normalized Haar measure [28, 29] on $U(d)$. The twirling map over $U(d)$, $\mathcal{T}$ is defined as the following:

$$\mathcal{T}(\mathcal{E}) := \int_{U(d)} (\mathcal{U}_{B_1}^{-1} \otimes \mathcal{U}_{B_2}^{-1}) \circ \mathcal{E} \circ \mathcal{U}_B d\mu_H(U) \quad (15)$$

where $\mathcal{E} : \mathcal{H}_B \to \mathcal{H}_{B_1} \otimes \mathcal{H}_{B_2}$ is a linear map, and $\mathcal{U}_B(\cdot) = U_B(\cdot)U_B^\dagger$ denotes the unitary conjugation map on subsystem $B$.

As a first step, we can observe that the sample complexity of a virtual broadcasting map remains unchanged under unitary twirling, giving us the following proposition:

**Proposition 3.** Recall that $\tilde{S}_\delta = \tilde{S}(\delta,\delta)$, where $\tilde{S}(\delta,\delta)$ denotes the square root of the sample complexity associated with the $(\delta,\delta)$-approximate broadcasting task. We define

$$\begin{aligned}
\tilde{S}_\delta^{\text{cov}} = \min \quad & x + y \\
\text{s.t.} \quad & \frac{1}{2} \|\text{Tr}_{B_2} \circ \mathcal{E} - \text{id}_{B \to B_1}\|_\diamond \leq \delta, \\
& \frac{1}{2} \|\text{Tr}_{B_1} \circ \mathcal{E} - \text{id}_{B \to B_2}\|_\diamond \leq \delta, \\
& (\mathcal{U} \otimes \mathcal{U}) \circ \mathcal{E} = \mathcal{E} \circ \mathcal{U} \text{ for all } U \in U(d) \\
& \mathcal{E} = \mathcal{E}_1 - \mathcal{E}_2, \\
& \text{Tr}_{B_1 B_2}[\mathcal{J}^{\mathcal{E}_1}] = xI_B, \\
& \text{Tr}_{B_1 B_2}[\mathcal{J}^{\mathcal{E}_2}] = yI_B, \\
& x - y = 1, \\
& \mathcal{J}^{\mathcal{E}_1} \geq 0, \quad \mathcal{J}^{\mathcal{E}_2} \geq 0.
\end{aligned} \quad (16)$$

Then $\tilde{S}_\delta^{\text{cov}} = \tilde{S}_\delta$. Here, $\mathcal{U}(\cdot) = U(\cdot)U^\dagger$ refers to the channel given by unitary evolution by the unitary $U$ from the $d$-dimensional unitary group $U(d)$.

*Proof.* We aim to show $\tilde{S}_\delta^{\text{cov}} = \tilde{S}_\delta$. It is clear that $\tilde{S}_\delta \leq \tilde{S}_\delta^{\text{cov}}$, since the set of feasible points of $\tilde{S}_\delta^{\text{cov}}$ form a subset of the feasible points for $\tilde{S}_\delta$. It remains to show $\tilde{S}_\delta^{\text{cov}} \leq \tilde{S}_\delta$. Take $(\mathcal{J}^{\mathcal{E}_1}, \mathcal{J}^{\mathcal{E}_2})$ be optimal solutions to the optimization problem associated with $\tilde{S}_\delta$. We claim that $(\mathcal{J}^{\mathcal{T}(\mathcal{E}_1)}, \mathcal{J}^{\mathcal{T}(\mathcal{E}_2)})$ is feasible for the optimization problem associated with $\tilde{S}_\delta^{\text{cov}}$. Thus establishing the upper bound.

Note that the twirling map $\mathcal{T}$ is a valid quantum superchannel. Therefore, it preserves both complete positivity and the trace-preserving property of the maps it acts upon. Since $\text{Tr}_{B_1 B_2}[\mathcal{J}^{\mathcal{E}_1}] = xI_B$, the scaled map $\frac{1}{x}\mathcal{E}_1$ is trace-preserving. Applying the superchannel, the map $\mathcal{T}(\frac{1}{x}\mathcal{E}_1)$ remains trace-preserving, which implies $\text{Tr}_{B_1 B_2}[\mathcal{J}^{\mathcal{T}(\frac{1}{x}\mathcal{E}_1)}] = I_B$, and thus $\text{Tr}_{B_1 B_2}[\mathcal{J}^{\mathcal{T}(\mathcal{E}_1)}] = xI_B$, satisfying the SDP constraint. Furthermore, because $\mathcal{E}_1$ is completely positive and $\mathcal{T}$ preserves complete positivity, $\mathcal{T}(\mathcal{E}_1)$ is completely positive, meaning its Choi operator is positive semi-definite. A similar argument holds for $\mathcal{J}^{\mathcal{E}_2}$.

It thus remains to check that the diamond norm constraints remain satisfied. Denoting by $\mathop{\mathbb{E}}\limits_{U \sim U(d)}$ the expectation value of an expression under the Haar distribution on $U(d)$, We have

$$\begin{aligned}
&\|\mathrm{Tr}_{B_1} \circ \mathcal{T}(\mathcal{E}) - \mathrm{id}_{B \to B_2}\|_\diamond \\
&= \left\| \mathop{\mathbb{E}}_{U \sim U(d)} \mathcal{U}^{-1} \circ \mathrm{Tr}_{B_1} \circ \mathcal{E} \circ \mathcal{U} - \mathop{\mathbb{E}}_{V \sim U(d)} \mathcal{V}^{-1} \circ \mathrm{id}_{B \to B_2} \circ \mathcal{V} \right\|_\diamond \\
&\leq \mathop{\mathbb{E}}_{U \sim U(d)} \left\| \mathcal{U}^{-1} \circ \mathrm{Tr}_{B_1} \circ \mathcal{E} \circ \mathcal{U} - \mathcal{U}^{-1} \circ \mathrm{id}_{B \to B_2} \circ \mathcal{U} \right\|_\diamond \\
&= \mathop{\mathbb{E}}_{U \sim U(d)} \left\| \mathcal{U}^{-1} \circ \mathrm{Tr}_{B_1} \circ (\mathcal{E} - \mathrm{id}_{B \to B_2}) \circ \mathcal{U} \right\|_\diamond \\
&= \mathop{\mathbb{E}}_{U \sim U(d)} \left\| \mathrm{Tr}_{B_1} \circ \mathcal{E} - \mathrm{id}_{B \to B_2} \right\|_\diamond \\
&= \left\| \mathrm{Tr}_{B_1} \circ \mathcal{E} - \mathrm{id}_{B \to B_2} \right\|_\diamond,
\end{aligned} \quad (17)$$

where the second inequality holds via the convexity, the third by linearity, the fourth by the unitary invariance of the diamond norm, and the final equality by the normalization of the Haar measure. Thus, the diamond norm constraints are satisfied and we are done. □

The above theorem demonstrates that one does not lose optimality in terms of sample complexity by restricting to unitary covariant broadcasting maps.

**Lemma 1** ([30]). *Let $\mathcal{E}_{B \to B_1 B_2}$ be an HPTP map. Suppose that $(\mathcal{U} \otimes \mathcal{U}) \circ \mathcal{E} = \mathcal{E} \circ \mathcal{U}$ for all $U \in U(d)$. Then on each marginal, the Choi operator of the twirled broadcasting map $\mathcal{T} \circ \mathrm{Tr}_{B_1} \circ \mathcal{E}$ is combination of $\Gamma_{BB_2}$ and $I_{BB_2}$, i.e.*

$$\mathcal{J}^{\mathcal{T} \circ \mathrm{Tr}_{B_1} \circ \mathcal{E}} = (1 - \lambda)\Gamma_{BB_2} + \lambda \frac{I}{d} \quad (18)$$

*for some $\lambda \in \mathbb{R}$.*

**Lemma 2.** *Let $\mathcal{R}(\cdot)$ be the replacement channel defined by*

$$\mathcal{R}(\rho) = \mathrm{Tr}[\rho]\left(\frac{I}{d}\right) \quad (19)$$

*then we have $\frac{1}{2}\|id - \mathcal{R}\|_\diamond = 1 - \frac{1}{d^2}$.*

**Lemma 3.** *Define the family of parameterized depolarizing noise channels by its Choi operator*

$$\Lambda^t = (1-t)\Gamma + t\frac{I}{d} \quad (20)$$

*for $t \in \mathbb{R}$. Note here since we allow for negative values of $t$, the resulting map may not necessarily be CPTP, but only HPTP. Let $Z : \mathbb{R} \to \mathbb{R}_{\geq 0}$ be defined by*

$$\begin{aligned}
Z(t) := \min \quad & x + y \\
\text{s.t.} \quad & \mathrm{Tr}_{B_1}[\mathcal{J}^\mathcal{E}] = \Lambda^t \\
& \mathrm{Tr}_{B_2}[\mathcal{J}^\mathcal{E}] = \Lambda^t \\
& \mathcal{J}^\mathcal{E} = \mathcal{J}^{\mathcal{E}_1} - \mathcal{J}^{\mathcal{E}_2}, \\
& \mathrm{Tr}_{B_1 B_2}[\mathcal{J}^{\mathcal{E}_1}] = xI_B, \quad \mathrm{Tr}_{B_1 B_2}[\mathcal{J}^{\mathcal{E}_2}] = yI_B, \\
& \mathcal{J}^{\mathcal{E}_1} \geq 0, \quad \mathcal{J}^{\mathcal{E}_2} \geq 0. \\
& x - y = 1
\end{aligned} \quad (21)$$

*then $Z(t)$ is convex. Furthermore the following hold:*

1. *If the points associated with $|t'| > 1$ is feasible, the points with $t = 1$ are feasible, and so $Z(1) \leq Z(t')$ for all $t'$ with $|t'| > 1$.*

2. $Z(|t|) \leq Z(-|t|)$ *for $t \in [-1, 1]$.*

3. $Z(t_1) \leq Z(t_2)$ *if $0 \leq t_2 \leq t_1 \leq 1$.*

*Thus, without of optimality for a fixed $\delta > 0$, we can reduce to considering values of $t$ where $t \in [0, 1]$.*

With the above lemmas in hand, we have the following theorem:

**Theorem 4.** *Let $\delta \in [0, 1]$ be a fixed error threshold on each marginal channel. We define $\tilde{S}_\delta^{dep}$ by the following SDP:*

$$\begin{aligned}
\tilde{S}_\delta^{dep} = \min \quad & x + y \\
\text{s.t.} \quad & \mathrm{Tr}_{B_1}[\mathcal{J}^\mathcal{E}] = \Lambda^t \\
& \mathrm{Tr}_{B_2}[\mathcal{J}^\mathcal{E}] = \Lambda^t \\
& \mathcal{J}^\mathcal{E} = \mathcal{J}^{\mathcal{E}_1} - \mathcal{J}^{\mathcal{E}_2}, \\
& \mathrm{Tr}_{B_1 B_2}[\mathcal{J}^{\mathcal{E}_1}] = xI_B, \quad \mathrm{Tr}_{B_1 B_2}[\mathcal{J}^{\mathcal{E}_2}] = yI_B, \\
& \mathcal{J}^{\mathcal{E}_1} \geq 0, \quad \mathcal{J}^{\mathcal{E}_2} \geq 0. \\
& x - y = 1 \\
& t = \frac{\delta d^2}{d^2 - 1}
\end{aligned} \quad (22)$$

*Then we have $\tilde{S}_\delta^{dep} = \tilde{S}_\delta$*

*Proof.* From our previous proposition, we have that the optimization can be restricted to the set of channels that are unitarily covariant. As in,

$$\tilde{S}_\delta^{\text{dep}} = \min \quad x + y$$
$$\text{s.t.} \quad \frac{1}{2} \left\| \text{Tr}_{B_2} \circ \mathcal{E} - \text{id}_{B \to B_1} \right\|_\diamond \leq \delta,$$
$$\frac{1}{2} \left\| \text{Tr}_{B_1} \circ \mathcal{E} - \text{id}_{B \to B_2} \right\|_\diamond \leq \delta,$$
$$(\mathcal{U} \otimes \mathcal{U}) \circ \mathcal{E} = \mathcal{E} \circ \mathcal{U} \text{ for all } U \in U(d)$$
$$\mathcal{E} = \mathcal{E}_1 - \mathcal{E}_2,$$
$$\text{Tr}_{B_1 B_2}[\mathcal{J}^{\mathcal{E}_1}] = x I_B,$$
$$\text{Tr}_{B_1 B_2}[\mathcal{J}^{\mathcal{E}_2}] = y I_B,$$
$$x - y = 1$$
$$\mathcal{J}^{\mathcal{E}_1} \geq 0, \quad \mathcal{J}^{\mathcal{E}_2} \geq 0.$$
(23)

The proof proceeds in a few stages. First, we claim that we can without loss of optimality restrict our optimization to HPTP maps whose marginal Choi operators are combinations of the parameterized depolarizing channels given above. Afterwards, we show that the requirement that $\frac{1}{2} \left\| \text{Tr}_{B_2} \circ \mathcal{E} - \text{id}_{B \to B_1} \right\|_\diamond \leq \delta$ is equivalent to the contraint that $|t| \leq \frac{\delta d^2}{d^2-1}$. Then we show that we can remove the absolute values in the constraints and restrict to the equality case without loss of optimality.

Let us first deal with the unitary covariance constraints. Now, take $(\mathcal{J}^{\mathcal{E}_1}, \mathcal{J}^{\mathcal{E}_2})$ feasible. Since $\mathcal{E} = \mathcal{E}_1 - \mathcal{E}_2$ satisfies unitary covariance, we have by lemma 1 that $\text{Tr}_{B_i}[\mathcal{J}^{\mathcal{E}}] = \Lambda^t$ for some $t \in \mathbb{R}$.

Secondly, we want to restrict our feasible set to values of $t$ such that

$$\frac{1}{2} \left\| \text{Tr}_{B_1} \circ \mathcal{E} - \text{id}_{B \to B_2} \right\|_\diamond \leq \delta \qquad (24)$$

To do this, observe that

$$\frac{1}{2} \left\| \text{Tr}_{B_1} \circ \mathcal{E} - \text{id}_{B \to B_2} \right\|_\diamond = \frac{1}{2} \left\| (1-t) id + t \mathcal{R} - id \right\|_\diamond$$
$$= |t| \frac{1}{2} \left\| \mathcal{R} - id \right\|_\diamond$$
$$= |t| \left( \frac{d^2 - 1}{d^2} \right)$$
(25)

where the first equality uses the fact that channels whose Choi operators are isotropic states are combinations of the identity and replacement channels, the second equality uses the homogeneity of scalars and the last equality uses lemma 2. Thus to satisfy the diamond norm constraints we require the choice of $|t| \leq \frac{\delta d^2}{d^2-1} \leq 1$. Next, lemma 3 allows us to without loss of optimality restrict to the cases where $t = \frac{\delta d^2}{d^2-1}$. Thus we have that $\tilde{S}_\delta^{\text{dep}} \leq \tilde{S}_\delta^{\text{cov}} = \tilde{S}_\delta$. However, it is clear that a point feasible for $\tilde{S}_\delta^{\text{dep}}$ is feasible for $\tilde{S}_\delta$ since restrictions on the parameter $t$ imply the norm constraints are satisfied by equation (25). Thus $\tilde{S}_\delta \leq \tilde{S}_\delta^{\text{dep}}$ and we have equality. □

Therefore, the fact that we can restrict our optimization to the set of channels that are unitary covariant results in a drastically simpler form for the SDP. This result also extends the work done in [18, 19] to describe the hierarchy of conditions that one might impose on a virtual broadcasting map. In particular, whilst [18] prove that there is only a single map satisfying the conditions of unitary covariance, classical consistency, and permutation invariance whose sample complexity overhead did not meet the SE condition, here we see that the class of unitary covariant approximate virtual broadcasting maps is rich enough to contain approximate virtual broadcasting maps that have optimal sample overhead, no matter the error thresholds on each marginal channel.

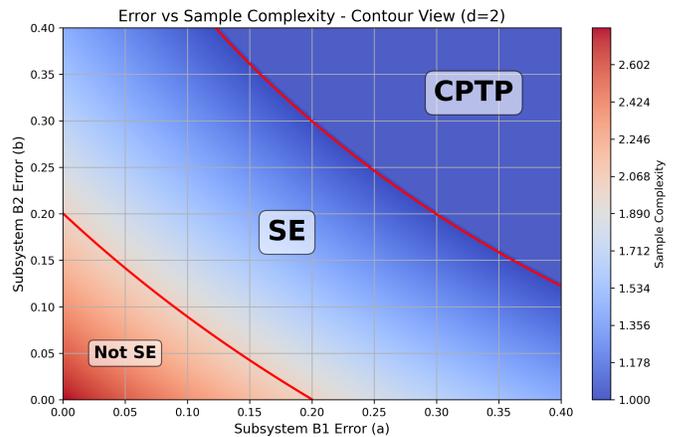

FIG. 3. **A contour plot of the map $S(a, b)$ that demonstrates for each pair of error thresholds $(a, b)$ the minimum sample complexity to implement the protocol on qubits.** The contour line for $\mathcal{S}(a,b) = 2$ and $\mathcal{S}(a,b) = 1$ are highlighted, demarcating the regions where it is possible for a broadcasting map to satisfy sample efficiency and when it can be implemented via a CPTP map. Similar plots for higher dimensions may be found in the appendix.

## V. TRADE-OFFS BETWEEN BIAS AND SAMPLE COMPLEXITY

In this section, we discuss the inverse to the problem considered in the previous section. That is, given a fixed budget for sample complexity, we ask what is the minimum error we can achieve using virtual approximate broadcast maps. This is formalized through the following:

**Definition 6.** Let $d \in \mathbb{N}$ be a the dimension $\mathcal{H}$, and let $\gamma \in \mathbb{R}_{\geq 1}$ be a fixed sample complexity budget. Then

$$\mu(\gamma, d) = \min_{(a,b) \in [0,1]^2} \{\max\{a,b\} : S(a,b) \leq \gamma\}$$
$$= \min_{\delta \in [0,1]} \{\delta : S(\delta) \leq \gamma\}$$
(26)

Where the second inequality follows from the fact that $S(a, b)$ is jointly convex and symmetric, and therefore the minimum value for a fixed $\gamma = a + b$ is achieved at the balanced point.

Therefore, by theorem 4 and proposition 2, we can derive the following SDP to compute $\mu(\gamma, d)$:

$$\begin{aligned}
\mu(\gamma, d) = \min \quad & \delta \\
\text{s.t.} \quad & \text{Tr}_{B_1}[\mathcal{J}^{\mathcal{E}}] = \Lambda^t \\
& \text{Tr}_{B_2}[\mathcal{J}^{\mathcal{E}}] = \Lambda^t \\
& t = \frac{\delta d^2}{d^2 - 1} \\
& \mathcal{J}^{\mathcal{E}} = \mathcal{J}^{\mathcal{E}_1} - \mathcal{J}^{\mathcal{E}_2}, \\
& \text{Tr}_{B_1 B_2}[\mathcal{J}^{\mathcal{E}_1}] = x I_B, \\
& \text{Tr}_{B_1 B_2}[\mathcal{J}^{\mathcal{E}_2}] = y I_B, \\
& \mathcal{J}^{\mathcal{E}_1} \geq 0, \quad \mathcal{J}^{\mathcal{E}_2} \geq 0. \\
& x - y = 1 \\
& (x + y)^2 \leq \gamma
\end{aligned} \quad (27)$$

The above SDP, along with the numerical data we obtain, quantifies precisely the amount of bias required to keep the sample overhead below any desired threshold, and in particular, to satisfy the SE condition.

Whilst it was shown in [19] that no exact virtual broadcasting map achieves the condition of sample efficiency, the above SDP allows us to show numerically that from dimensions 2 through 10, injecting a small amount of bias allows one to construct a broadcasting map that satisfies SE. In particular, $\mu(\gamma, d)$ tells us the error on each marginal associated to the optimal broadcasting map, if we restrict only to HPTP maps with sample complexity below $\gamma$. For reference, minimum magnitude of such bias, when we restrict to maps with sample complexity below 1.8, is displayed in figure 4 for dimensions 2 through 10.

The following question remains: for arbitrary dimensions, can we always find an approximate broadcasting protocol that achieves the SE condition? If so, how much bias do we have to tolerate? To answer this question, we give an upper bound for the value of $\mu(\gamma, d)$, which shows that approximate broadcasting maps that satisfy the condition of sample efficiency always exist, regardless of dimension.

**Theorem 5.** *For all $d \geq 2$, we have the upper bound $\mu(\gamma, d) \leq \frac{d^2-1}{d^2}(\frac{3-\sqrt{\gamma}}{4})$. Which gives us the upper bound $\mu(\gamma, d) \leq (\frac{3-\sqrt{\gamma}}{4})$ independent of the dimension of the subsystems.*

This result is proven by giving an explicit feasible point in the SDP. In particular, we let

$$\begin{aligned}
\mathcal{J}^{\mathcal{E}_1} &= \frac{\sqrt{\gamma}+1}{4} \left( \frac{\Gamma_{BB_1} \otimes I_{B_2}}{d} + \frac{\Gamma_{BB_2} \otimes I_{B_1}}{d} \right) \\
\mathcal{J}^{\mathcal{E}_2} &= \frac{\sqrt{\gamma}-1}{2} \left( \frac{\Gamma_{B_1 B_2} \otimes I_B}{d} \right)
\end{aligned} \quad (28)$$

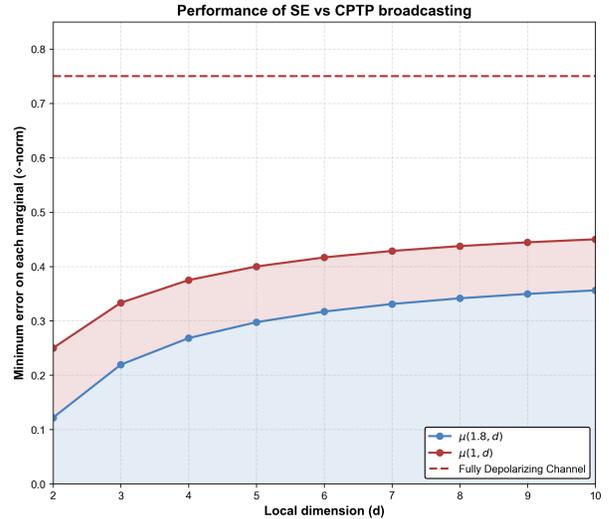

FIG. 4. **Comparison of the minimum diamond norm error on each marginal channel for the case of approximate virtual broadcasting in a regime satisfying the SE condition, and the minimum diamond norm error if one restricts to only using quantum channels without Monte-Carlo sampling**. We show numerically that for dimensions 2 through 10, virtual operations outperform quantum channels for the task of approximate broadcasting. In particular, for the qubit case, the diamond norm error is $\approx 0.12$, as compared to the error of 0.25 one obtains when restricted to quantum channels, and the maximum error of 0.75 one obtains by using the fully depolarizing channel.

and check that they satisfy the SDP corresponding to $\mu(2, d)$. This gives us upper bound of $\frac{d^2-1}{d^2}(\frac{3-\sqrt{\gamma}}{4})$.

*Proof.* We show this bound by giving a feasible point in the minimization problem. Let

$$\begin{aligned}
\mathcal{J}^{\mathcal{E}_1} &= \frac{\sqrt{\gamma}+1}{4} \left( \frac{\Gamma_{BB_1} \otimes I_{B_2}}{d} + \frac{\Gamma_{BB_2} \otimes I_{B_1}}{d} \right) \\
\mathcal{J}^{\mathcal{E}_2} &= \frac{\sqrt{\gamma}-1}{2} \left( \frac{\Gamma_{B_1 B_2} \otimes I_B}{d} \right)
\end{aligned} \quad (29)$$

Both Choi operators are manifestly positive-semidefinite. Furthermore, we have

$$\begin{aligned}
\text{Tr}_{B_1 B_2}[\mathcal{J}^{\mathcal{E}_1}] &= \frac{\sqrt{\gamma}+1}{4}(I_B + I_B) = \frac{\sqrt{\gamma}+1}{2} I_B \\
\text{Tr}_{B_1 B_2}[\mathcal{J}^{\mathcal{E}_2}] &= \frac{\sqrt{\gamma}-1}{2} I_B
\end{aligned} \quad (30)$$

So if we let $x = \frac{\sqrt{\gamma}+1}{2}$ and $y = \frac{\sqrt{\gamma}-1}{2}$, we have $(x+y)^2 = 2$ and $x - y = 1$. Now we check that this term reduces to a parameterized depolarizing noise channel on each

marginal. We have

$$\text{Tr}_{B_1}[\mathcal{J}^{\mathcal{E}}] = \text{Tr}_{B_1}[\mathcal{J}^{\mathcal{E}_1}] - \text{Tr}_{B_1}[\mathcal{J}^{\mathcal{E}_2}]$$
$$= \frac{\sqrt{\gamma}+1}{4}\left(\frac{I_{BB_2}}{d} + \Gamma_{BB_2}\right) - \frac{\sqrt{\gamma}-1}{2}\frac{I_{BB_2}}{d}$$
$$= \left(1 - \frac{3-\sqrt{\gamma}}{4}\right)\Gamma_{BB_2} + \left(\frac{3-\sqrt{\gamma}}{4}\right)\frac{I_{BB_2}}{d} \tag{31}$$

and similarly

$$\text{Tr}_{B_2}[\mathcal{J}^{\mathcal{E}}] = \left(1 - \frac{3-\sqrt{\gamma}}{4}\right)\Gamma_{BB_1} + \left(\frac{3-\sqrt{\gamma}}{4}\right)\frac{I_{BB_1}}{d} \tag{32}$$

Thus, if we pick $\delta = \frac{d^2-1}{d^2}(\frac{3-\sqrt{\gamma}}{4})$, the point $(\delta, \mathcal{J}^{\mathcal{E}_1}, \mathcal{J}^{\mathcal{E}_2})$ is feasible and we get the upper bound $\mu(\gamma, d) \leq \frac{d^2-1}{d^2}(\frac{3-\sqrt{\gamma}}{4})$. $\square$

Furthermore, the constructive proof shows that our protocol can be implemented by sampling from combinations of two very elementary "discard-and-prepare" channels:

- $\frac{\Gamma_{BB_1} \otimes I_{B_2}}{d}$. We teleport the input system $B$ into one receiver, $B_1$ (or $B_2$), and replace the other output by the maximally mixed state.

- $\frac{\Gamma_{B_1B_2} \otimes I_B}{d}$. We discard $B$ entirely and prepare the maximally mixed state on both receiving subsystems.

By sampling combinations of these two simple maps, which are easy to characterize, we obtain the required approximate virtual broadcasting channel. Thus the construction is explicit, and does not involve any exotic terms.

Whilst the construction given above is not unitary covariant, by theorem 4, there always exists a corresponding protocol that achieves the equal sample overhead that is unitary covariant. Therefore, regardless of dimension, one can always construct a 1-to-2 approximate broadcasting protocol that satisfies SE and unitary covariance.

The above result answers the one of the questions we began with in this work. Namely, we demonstrate the existence of approximate virtual broadcasting protocols that achieve both the SE condition and are unitarily covariant. Furthermore, since the marginal errors remain bounded independent of dimension, the above result shows that approximate virtual broadcasting is achievable in a genuinely non-trivial way. Not only is the diamond norm error small for qubits ($\approx 0.1$), but in fact it is capped by a universal constant ($\approx 0.42$), independent of the dimensions of the subsystems involved!

In [31], the optimal fidelity of quantum broadcasting with a single quantum channel was characterized through an SDP. With our SDP formulation of $\mu(\gamma, d)$, we demonstrate that our proposed framework of approximate virtual broadcasting has an advantage over approximate broadcasting using a quantum channel when the sample overhead is restricted to be below that required for naive sample-splitting. Thus, approximate virtual broadcasting not only achieves the desired sample efficiency condition but also serves as an example of how virtual operations can outperform physical quantum channels. To compare the performance of an approximate virtual broadcasting protocol against approximate virtual broadcasting using a quantum channel, it suffices to compare $\mu(1.8, d)$ and $\mu(1, d)$. We show in figure 4, numerically, that $\mu(1, d) \geq \mu(1.8, d)$ for dimensions 2 through 10. This demonstrates a robust advantage compared to simply using a quantum channel, not just in the qubit case, but one that appears to scale well into higher dimensions.

## VI. CONCLUSIONS AND FUTURE DIRECTIONS

In this work, we examine the possibility of constructing an approximate virtual quantum broadcasting map satisfying the condition of sample efficiency. We prove analytically that such a map exists and numerically determine the minimal bias required to satisfy the sample-efficiency condition. Furthermore, we show that its deviation from exact broadcasting (quantified by diamond-norm distance) remains uniformly bounded in arbitrary dimensions. Finally, we show that approximate virtual broadcasting maps satisfying the sample-efficiency condition are still able to outperform quantum channels in terms of diamond-norm error. Therefore, our work highlights that efficient quantum broadcasting is not only of conceptual interest but can also be implemented as a simple Monte Carlo protocol with meaningful operational advantages.

Several questions remain open. While we explored the trade-off between diamond-norm error and sample complexity, the trade-off between error and other resources such as entanglement remains unexplored. In the scenario where our aim is to construct a sample-efficient and unitary-covariant broadcasting map, one of the channels that we would need to implement is the identity channel of systems $B \to B_i$; this requires entanglement. If one instead optimizes for minimal entanglement cost, does the optimal broadcasting map change? Can such a map still satisfy unitary covariance [18] and sample efficiency simultaneously [19, 20], and what additional structural properties would emerge? Another promising direction is to combine virtual broadcasting with virtual resource distillation [11–13]: how does such a hybrid protocol perform, and can it reduce auxiliary costs or amplify the effective advantage? Addressing these questions would both tighten the resource-theoretic limits of virtual operations and shed light on more general constraints in quantum information processing.

ACKNOWLEDGEMENTS

The authors thank Foo Yan Xi for help in running part of the numerical computations, and Ximing Wang and Yunlong Xiao for extensive discussions during the early stages of this work. At the time of its completion, we became aware of Wang and Xiao's independent work on the same topic, titled "Practical Quantum Broadcasting". MST acknowledges support from A*STAR via the A*STAR Research Internship Award (ARIA).

[1] Peter W Shor. Algorithms for quantum computation: discrete logarithms and factoring. In *Proceedings 35th annual symposium on foundations of computer science*, pages 124–134. Ieee, 1994.

[2] Lov K Grover. A fast quantum mechanical algorithm for database search. In *Proceedings of the twenty-eighth annual ACM symposium on Theory of computing*, pages 212–219, 1996.

[3] Charles H Bennett and Stephen J Wiesner. Communication via one-and two-particle operators on einstein-podolsky-rosen states. *Physical review letters*, 69(20):2881, 1992.

[4] Charles H Bennett, Gilles Brassard, Claude Crépeau, Richard Jozsa, Asher Peres, and William K Wootters. Teleporting an unknown quantum state via dual classical and einstein-podolsky-rosen channels. *Physical review letters*, 70(13):1895, 1993.

[5] William K Wootters and Wojciech H Zurek. A single quantum cannot be cloned. *Nature*, 299(5886):802–803, 1982.

[6] Giulio Chiribella. On quantum estimation, quantum cloning and finite quantum de finetti theorems. In *Conference on Quantum Computation, Communication, and Cryptography*, pages 9–25. Springer, 2010.

[7] Howard Barnum, Jonathan Barrett, Matthew Leifer, and Alexander Wilce. Generalized no-broadcasting theorem. *Physical review letters*, 99(24):240501, 2007.

[8] Howard Barnum, Carlton M Caves, Christopher A Fuchs, Richard Jozsa, and Benjamin Schumacher. Noncommuting mixed states cannot be broadcast. *Physical Review Letters*, 76(15):2818, 1996.

[9] Jiaqing Jiang, Kun Wang, and Xin Wang. Physical implementability of linear maps and its application in error mitigation. *Quantum*, 5:600, 2021.

[10] Francesco Buscemi, Michele Dall'Arno, Masanao Ozawa, and Vlatko Vedral. Direct observation of any two-point quantum correlation function. *arXiv preprint arXiv:1312.4240*, 2013.

[11] Xiao Yuan, Bartosz Regula, Ryuji Takagi, and Mile Gu. Virtual quantum resource distillation. *Physical Review Letters*, 132(5):050203, 2024.

[12] Ting Zhang, Yukun Zhang, Lu Liu, Xiao-Xu Fang, Qian-Xi Zhang, Xiao Yuan, and He Lu. Experimental virtual distillation of entanglement and coherence. *Physical Review Letters*, 132(18):180201, 2024.

[13] Ryuji Takagi, Xiao Yuan, Bartosz Regula, and Mile Gu. Virtual quantum resource distillation: General framework and applications. *Physical Review A*, 109(2):022403, 2024.

[14] Zhi-Hao Bi, Jing-Tao Qiu, and Xiao-Dong Yu. Virtual cloning of quantum states. *Physical Review A*, 112(1):012420, 2025.

[15] Xuanqiang Zhao, Xin Wang, and Giulio Chiribella. Shadow simulation of quantum processes. *Physical Review Letters*, 133(12):120804, 2024.

[16] Chengkai Zhu, Yin Mo, Yu-Ao Chen, and Xin Wang. Reversing unknown quantum processes via virtual combs for channels with limited information. *Physical Review Letters*, 133(3):030801, 2024.

[17] Hongshun Yao, Xia Liu, Chengkai Zhu, and Xin Wang. Optimal unilocal virtual quantum broadcasting. *Physical Review A*, 110(1):012458, 2024.

[18] Arthur J Parzygnat, James Fullwood, Francesco Buscemi, and Giulio Chiribella. Virtual quantum broadcasting. *Physical Review Letters*, 132(11):110203, 2024.

[19] Yunlong Xiao, Xiangjing Liu, and Zhenhuan Liu. No practical quantum broadcasting: Even virtually. *Physical Review Letters*, 135(9):090202, 2025.

[20] Yunlong Xiao, Xiangjing Liu, and Zhenhuan Liu. No practical quantum broadcasting: General framework. *Physical Review Research*, 7(3):033194, 2025.

[21] Reiji Okada and Francesco Buscemi. Virtual phase-covariant quantum broadcasting for qubits. *arXiv preprint arXiv:2511.20014*, 2025.

[22] Lieven Vandenberghe and Stephen Boyd. Semidefinite programming. *SIAM review*, 38(1):49–95, 1996.

[23] Lieven Vandenberghe and Stephen Boyd. Applications of semidefinite programming. *Applied Numerical Mathematics*, 29(3):283–299, 1999.

[24] Paul Skrzypczyk and Daniel Cavalcanti. *Semidefinite programming in quantum information science*. IOP Publishing, 2023.

[25] Man-Duen Choi. Completely positive linear maps on complex matrices. *Linear algebra and its applications*, 10(3):285–290, 1975.

[26] Andrzej Jamiołkowski. Linear transformations which preserve trace and positive semidefiniteness of operators. *Reports on mathematical physics*, 3(4):275–278, 1972.

[27] Wassily Hoeffding. Probability inequalities for sums of bounded random variables. *Journal of the American statistical association*, 58(301):13–30, 1963.

[28] Antonio Anna Mele. Introduction to haar measure tools in quantum information: A beginner's tutorial. *Quantum*, 8:1340, 2024.

[29] Benoît Collins and Piotr Śniady. Integration with respect to the haar measure on unitary, orthogonal and symplectic group. *Communications in Mathematical Physics*, 264(3):773–795, 2006.

[30] Michal Horodecki and Pawel Horodecki. Reduction criterion of separability and limits for a class of protocols of entanglement distillation, 1998.

[31] Wei Xie, Kun Fang, Xin Wang, and Runyao Duan. Approximate broadcasting of quantum correlations. *physical Review A*, 96(2):022302, 2017.

[32] Sumeet Khatri and Mark M Wilde. Principles of quantum communication theory: A modern approach. *arXiv preprint arXiv:2011.04672*, 2020.

# VII. APPENDIX

## A. QUANTUM CHANNELS AND VIRTUAL QUANTUM BROADCASTING MAPS

In this work, we restrict to the case of broadcasting from one source to two clients. We will call the source $B$ and the clients $B_1$ and $B_2$. Each of these systems is a qudit, i.e., associated with a Hilbert space $\mathcal{H}$ of finite dimension $d$. We denote by $\mathcal{B}(\mathcal{H})$ the space of bounded linear operators acting on $\mathcal{H}$. Quantum states of a qubit are associated with density operators $\rho \in \mathcal{B}(\mathcal{H})$, satisfying $\rho \geq 0$ and $\text{Tr}[\rho] = 1$; denote the space of all such density operators by $\mathcal{D}(\mathcal{H})$. Qudit-to-qudit quantum channels are associated with linear endomorphisms of $\mathcal{B}(\mathcal{H})$ that are completely positive (CP) and trace-preserving (TP). We fix a canonical orthonormal basis $\{|i\rangle\}$ for $\mathcal{H}$, and define $\Gamma := \sum_i |ii\rangle \in \mathcal{H} \otimes \mathcal{H}$. For a channel $\mathcal{E}$, its Choi operator (relative to this basis) is $\mathcal{J}^{\mathcal{E}} := (I \otimes \mathcal{E})\Gamma$. Finally, we will omit qudit labels on operator and superoperator symbols where the associated qudits are clear from the context. When we do use explicit labels, we adopt the following convention: if $L$ is a one-qudit linear operator, then $L_B$ will denote $L$ instantiated on the space of system $B$; similarly, for a two-qudit operator, $M_{BB_1}$ will denote $M$ instantiated on the spaces of $BB_1$, etc. In such cases, we will not use the convention where $M_B$ denotes the corresponding marginal, instead denoting the latter explicitly as $\text{Tr}_{B_1} M_{BB_1}$, or simply $\text{Tr}_{B_1} M$. Choi operators that represent quantum channels are characterized by the following proposition:

**Proposition 4.** *Let $\mathcal{E}$ be a linear map from systems $A$ to $B$ and $\mathcal{J}^{\mathcal{E}}$ be its associated Choi operator. Then $\mathcal{E}$ is a quantum channel if and only if its Choi operator $J^{\mathcal{E}} \geq 0$ and $\text{Tr}_B[\mathcal{J}^{\mathcal{E}}] = I_A$.*

Importantly, for the case of Hermitian preserving maps that we deal with in this work, their Choi operators can be characterized by the following proposition:

**Proposition 5.** *Let $\mathcal{E}$ be a linear map and $\mathcal{J}^{\mathcal{E}}$ be its associated Choi operator. Then $\mathcal{E}$ is Hermitian preserving if and only if its Choi operator $J^{\mathcal{E}}$ is Hermitian.*

To formalize the notion of broadcasting, we can define the notion of a broadcasting map.

**Definition 7** (Broadcasting map). $\mathcal{H}_B, \mathcal{H}_{B_1}$ and $\mathcal{H}_{B_2}$ be Hilbert spaces. A broadcasting map is then a Hermitian-preserving, trace-preserving (HPTP) map $\mathcal{E}_{B \to B_1 B_2}$ satisfying

$$\text{Tr}_{B_1}[\mathcal{E}(\rho)] = \rho \text{ and} \tag{33}$$
$$\text{Tr}_{B_2}[\mathcal{E}(\rho)] = \rho \tag{34}$$

for all $\rho \in \mathcal{D}(\mathcal{H})$.

Alternatively, this condition can be phrased in the language of Choi operators. Before that, we introduce the link product between two Hermitian operators, a useful operation that helps us simplify the Choi operators of multiple composed channels.

**Definition 8.** Let $A$ and $B$ be Hermitian operators acting on multiple subsystems, and $X$ being the common subsystems on whch $A$ and $B$ act, we can define the link product $\star$ between $A$ and $B$ by

$$A \star B := \text{Tr}_X[A^{T_X} B] \tag{35}$$

where $T_X$ denotes the partial transpose over system $X$.

A consequence of this is that $\mathcal{J}^{\mathcal{F} \circ \mathcal{E}} = \mathcal{J}^{\mathcal{F}} \star \mathcal{J}^{\mathcal{E}}$. The following proposition was given by [17], but here we offer a different proof we believe may be simpler for some to follow.

**Proposition 6** ([17]). *We say that $\mathcal{J}^{\mathcal{E}}$ is a Choi operator of a broadcast map if*

$$\text{Tr}_{B_2}[\mathcal{J}^{\mathcal{E}}] = \Gamma_{BB1} \text{ and} \tag{36}$$
$$\text{Tr}_{B_1}[\mathcal{J}^{\mathcal{E}}] = \Gamma_{BB2} \tag{37}$$

*Proof.* First suppose that $\mathcal{E}$ is a broadcasting map, then we know that $\text{Tr}_{B_1} \circ \mathcal{E} = id_{B \to B_2}$. Applying to Choi–Jamiołkowski isomorphism on both sides, we get

$$\mathcal{J}^{\text{Tr}_{B_1} \circ \mathcal{E}} = \mathcal{J}^{d_{B \to B_2}} \tag{38}$$

Since the Choi operator of the identity map is $\Gamma_{BB_2}$, and $\mathcal{J}^{\text{Tr}_{B_1} \circ \mathcal{E}} = I_{B_1} \star \mathcal{J}^{\mathcal{E}} = \text{Tr}_{B_1}[\mathcal{E}]$. Therefore we have

$$\text{Tr}_{B_1}[J^{\mathcal{E}}] = \Gamma_{BB_2} \tag{39}$$

A similar argument holds when we trace out the subsystem $B_2$.
For the other direction, the proof follows from applying the inverse of the Choi–Jamiołkowski isomorphism. □

Whilst it is shown that there are no universal broadcasting maps that satisfy the CPTP condition, examples of HPTP broadcasting maps abound. For example the one parameter family of HPTP maps

$$\rho \mapsto \frac{1}{2}\{\rho \otimes I, SWAP\} + i\lambda[\rho \otimes I, SWAP]$$

where $\lambda \in \mathbb{C}$ is a family of broadcasting channels satisfying desirable properties such as unitary covariance, and classical consistency (here $\{\cdot,\cdot\}, [\cdot,\cdot]$ denote the anti-commutator and commutator respectively). A particular important example is a broadcasting map satisfying the following set of conditions:
For a broadcasting map $\mathcal{E}$ and orthonormal basis $\{|i\rangle\}$:

1. **Unitary Covariance**
   $\mathcal{E} \circ \mathcal{U} = \mathcal{U}^{\otimes 2} \circ \mathcal{E}$ for any unitary gate $\mathcal{U}$.

2. **Permutation Invariance**
   $(SWAP)(\mathcal{E}(\rho))(SWAP) = \mathcal{E}(\rho)$, where $SWAP$ is the operator $\sum_{ij} |i\rangle\langle j| \otimes |j\rangle\langle i|$.

3. **Classical Consistency**
   $((\Delta \otimes \Delta) \circ \mathcal{E} \circ \Delta)(|i\rangle\langle j|) = \delta_{ij} |i\rangle\langle i| \otimes |i\rangle\langle i|$.

It was shown in [18] that the map

$$\rho \mapsto \frac{1}{2}\{\rho \otimes I, SWAP\} \tag{40}$$

is the unique map that satisfies the above three conditions, with the broadcasting and HPTP condition being satisified as a consequence of imposing the above conditions.

## B. Proofs of additional lemmata

In this section, we discuss some facts related to the diamond norm and its equivalent SDP formulations, give the proof of the three lemmas stated in section IV.

**Definition 9** (Diamond Norm). Let $H$ be a Hilbert space, and let $\mathcal{B}(\mathcal{H})$ be the set of linear operators acting on that Hilbert space. Let $\|\cdot\|_p$ denote the Schatten p-norms of the bounded operators on that space. Now let $\Phi$ be an HPTP linear operator on $\mathcal{B}(\mathcal{H})$. Then its diamond norm is defined by

$$\|\Phi\|_\diamond := \sup_{\|X\|_1 \leq 1} \|(\Phi \otimes I)(X)\|_1$$

where the X's are operators on the Hilbert space $\mathcal{H}^{\otimes 2}$. We should remark that in principle the diamond norm is defined by taking the supremum $(\Phi \otimes I_d)$ over all dimensions $d$, but this is achieved where $d$ is the same dimension as the system that $\Phi$ acts on.

The diamond norm can be computed efficiently using the following SDP formulation.

**Proposition 7** ([32]). *Let $\mathcal{E}_1$ and $\mathcal{E}_2$ be CPTP maps from a system $B$ to $B_1$. Then finding $\frac{1}{2}\|\mathcal{E}_1 - \mathcal{E}_2\|_\diamond$ is equivalent to the following SDP:*

$$\begin{aligned}
\min \quad & \mu \\
\text{s.t.} \quad & Z_{BB_1} \geq 0, \\
& Z_{BB_1} \geq \mathcal{J}^{\mathcal{E}_1} - \mathcal{J}^{\mathcal{E}_2}, \\
& \mu I_B \geq \text{Tr}_{B_1}[Z_{BB_1}],
\end{aligned} \quad (41)$$

In this section, we give the proof of the three lemmas stated in section IV
The first is the proof of lemma 1.
We give the proof of lemma 2.

**Lemma 6.** *Let $\mathcal{R}(\cdot)$ be the replacement channel defined by*

$$\mathcal{R}(\rho) = \text{Tr}[\rho]\left(\frac{I}{d}\right) \quad (42)$$

*then we have $\frac{1}{2}\|id - \mathcal{R}\|_\diamond = 1 - \frac{1}{d^2}$.*

*Proof.* Recall that the SDP to compute the diamond metric of $\frac{1}{2}\|id - \mathcal{R}\|_\diamond$ is given by

$$\begin{aligned}
\min \quad & \mu \\
\text{s.t.} \quad & 0 \leq \mu, \\
& Z_{BB_1} \geq \Gamma_{BB_1} - \frac{I_{BB_1}}{d}, \\
& \mu I \geq \text{Tr}_{B_1}[Z_{BB_1}].
\end{aligned}$$

We can use this to obtain a tight upper bound for the diamond norm of $id - \mathcal{R}$. Observe that we can simultaneously diagonalize both operators since they commute, and recall that $\Gamma_{BB_1}$ is a projector onto the one dimensional subspace spanned by $|\Gamma\rangle$ with eigenvalue $d$. So the spectrum of $\Gamma_{BB_1} - \frac{I}{d}$ is given by $\{d - \frac{1}{d}, \frac{-1}{d}, ..., \frac{-1}{d}\}$. So we can just choose $Z_{BB_1}$ to be $d - \frac{1}{d}(\frac{\Gamma}{d})$, so that $\text{Tr}_{B_1}[Z_{BB_1}] = (d - \frac{1}{d})\text{Tr}\left[\frac{\Gamma_{BB_1}}{d}\right] = (1 - \frac{1}{d^2})I$, and so we know that the upper bound for $\mu$ is $1 - \frac{1}{d^2}$, and thus we have $\frac{1}{2}\|id - \mathcal{R}\|_\diamond \leq 1 - \frac{1}{d^2}$.

To show that the lower bound is tight, recall that the for a hermitian operator, the trace norm is just the $\ell_1$ norm of the sequence of eigenvalues, so

$$\begin{aligned}
\|id - \mathcal{R}\|_\diamond &\geq \left\|((id - R) \otimes id)(\frac{\Gamma}{d})\right\|_1 \\
&= \frac{1}{d}\left\|\Gamma - \frac{I}{d}\right\|_1 \\
&= \frac{1}{d}(d - \frac{1}{d} + \frac{d^2 - 1}{d}) \\
&= 2 - \frac{2}{d^2}
\end{aligned} \quad (43)$$

Therefore we have $\frac{1}{2}\|id - \mathcal{R}\|_\diamond \geq 1 - \frac{1}{d^2}$ □

The following lemma leverages the isotropic twirling technique introduced in [30].

**Lemma 7.** *Let $\mathcal{E}_{B \to B_1 B_2}$ be an HPTP map. Suppose that $(\mathcal{U} \otimes \mathcal{U}) \circ \mathcal{E} = \mathcal{E} \circ \mathcal{U}$ for all $U \in U(d)$. Then on each marginal, the Choi operator of the twirled broadcasting map $\mathcal{T} \circ \mathrm{Tr}_{B_1} \circ \mathcal{E}$ is combination of $\Gamma_{BB_2}$ and $I_{BB_2}$, i.e.*

$$\mathcal{J}^{\mathcal{T} \circ \mathrm{Tr}_{B_1} \circ \mathcal{E}} = (1-\lambda)\Gamma_{BB_2} + \lambda \frac{I}{d} \tag{44}$$

*for some $\lambda \in \mathbb{R}$.*

*Proof.* This is because the 1-to-2 channel twirling is equivalent to isotropic twirling of the Choi operator. Consider the twirled broadcasting map on the marginal $\mathcal{T} \circ \mathrm{Tr}_{B_1} \circ \mathcal{E}$, defined by

$$\mathcal{T} \circ \mathrm{Tr}_{B_1} \circ \mathcal{E} = \int_{U(d)} \mathcal{U}^\dagger \circ \mathrm{Tr}_{B_1} \mathcal{E} \circ \mathcal{U} dU \tag{45}$$

where $\mathcal{U} = U(\cdot)U^\dagger$ is the map defined by unitary conjugation. Since the Choi isomorphism is linear, we get

$$\begin{aligned}
\mathcal{J}^{\mathcal{T} \circ \mathrm{Tr}_{B_1} \circ \mathcal{E}} &= \int_{U(d)} \mathcal{J}^{\mathcal{U}^\dagger \circ \mathrm{Tr}_{B_1} \circ \mathcal{U}} dU \\
&= \int_{U(d)} \mathcal{J}^{\mathcal{U}^\dagger} \star \mathcal{J}^{\mathrm{Tr}_{B_1} \circ \mathcal{E}} \star J^{\mathcal{U}} dU \\
&= \int_{U(d)} (U \otimes \overline{U})^\dagger J^{\mathrm{Tr}_{B_1} \circ \mathcal{E}} (U \otimes \overline{U}) dU
\end{aligned} \tag{46}$$

Since the map $U \mapsto U \otimes \overline{U}$ defines a representation of the unitary group $U(d)$ with the projectors onto the irreps being $\frac{\Gamma_{BB_2}}{d}$ with dimension 1 and $I - \frac{\Gamma_{BB_2}}{d}$ with dimension $d^2 - 1$ respectively, we have by Schur's lemma that:

$$\begin{aligned}
\mathcal{J}^{\mathcal{T} \circ \mathrm{Tr}_{B_1} \circ \mathcal{E}} &= \int_{U(d)} (U \otimes \overline{U})^\dagger J^{\mathrm{Tr}_{B_1} \circ \mathcal{E}} (U \otimes \overline{U}) dU \\
&= \frac{1}{d} \mathrm{Tr}\big[\Gamma_{BB_2} \mathcal{J}^{\mathrm{Tr}_{B_1} \circ \mathcal{E}}\big] \frac{\Gamma_{BB_2}}{d} + \frac{\mathrm{Tr}\big(I - \frac{1}{d}\Gamma_{BB_2}\big) \mathcal{J}^{\mathrm{Tr}_{B_1} \circ \mathcal{E}}\big]}{d^2 - 1}(I - \frac{\Gamma_{BB_2}}{d})
\end{aligned} \tag{47}$$

Now denote $\frac{1}{d} \mathrm{Tr}\big[\Gamma_{BB_2} \mathcal{J}^{\mathrm{Tr}_{B_1} \circ \mathcal{E}}\big] = F$ and the normalized maximum entangled state as $\Omega$, then we obtain

$$\begin{aligned}
\mathcal{J}^{\mathcal{T} \circ \mathrm{Tr}_{B_1} \circ \mathcal{E}} &= F\Omega + \frac{\mathrm{Tr}(1-\Omega)\mathcal{J}^{\mathrm{Tr}_{B_1} \circ \mathcal{E}}]}{d^2 - 1}(I - \Omega) \\
&= F\Omega + \frac{d - F}{d^2 - 1}(I - \Omega) \\
&= \frac{Fd - 1}{d^2 - 1}\Gamma_{BB_2} + \frac{d^2 - Fd}{d^2 - 1}\big(\frac{I}{d}\big)
\end{aligned} \tag{48}$$

Since $\mathcal{E}$ is HP, $\mathcal{J}^{\mathcal{E}}$ is Hermitian, thus $F \in \mathbb{R}$.

$\square$

Now we give the proof of 3.

**Lemma 8.** *Define the family of parameterized depolarizing noise channels by its Choi operator*

$$\Lambda^t = (1-t)\Gamma + t\frac{I}{d} \tag{49}$$

*for $t \in \mathbb{R}$. Note here since we allow for negative values of $t$, the resulting map may not necessarily be CPTP, but only HPTP. Let $Z: \mathbb{R} \to \mathbb{R}_{\geq 0}$ be defined by*

$$\begin{aligned}
Z(t) := \min \quad & x + y \\
\text{s.t.} \quad & \mathrm{Tr}_{B_1}[\mathcal{J}^{\mathcal{E}}] = \Lambda^t \\
& \mathrm{Tr}_{B_2}[\mathcal{J}^{\mathcal{E}}] = \Lambda^t \\
& \mathcal{J}^{\mathcal{E}} = \mathcal{J}^{\mathcal{E}_1} - \mathcal{J}^{\mathcal{E}_2}, \\
& \mathrm{Tr}_{B_1 B_2}[\mathcal{J}^{\mathcal{E}_1}] = xI_B, \quad \mathrm{Tr}_{B_1 B_2}[\mathcal{J}^{\mathcal{E}_2}] = yI_B, \\
& \mathcal{J}^{\mathcal{E}_1} \geq 0, \quad \mathcal{J}^{\mathcal{E}_2} \geq 0. \\
& x - y = 1
\end{aligned} \tag{50}$$

then $Z(t)$ is convex.

1. If the points associated with $|t'| > 1$ is feasible, the points with $t = 1$ are feasible, and so $Z(1) \leq Z(t')$ for all $t'$ with $|t'| > 1$.

2. $Z(|t|) \leq Z(-|t|)$ for $t \in [-1, 1]$.

3. $Z(t_1) \leq Z(t_2)$ if $0 \leq t_2 \leq t_1 \leq 1$.

Thus, without of optimality for a fixed $\delta > 0$, we can reduce to considering the set of $t \in [0, 1]$.

*Proof.* First, we prove convexity. Take $\bar{t}, \tilde{t}$ with solutions $(\mathcal{J}^{\overline{\mathcal{E}_1}}, \mathcal{J}^{\overline{\mathcal{E}_2}}, \overline{x}, \overline{y})$ and $(\mathcal{J}^{\tilde{\mathcal{E}_1}}, \mathcal{J}^{\tilde{\mathcal{E}_2}}, \tilde{x}, \tilde{y})$. Now fix $\lambda \in [0, 1]$ and take $t = \lambda \bar{t} + \lambda \tilde{t}$. Then we aim to upper bound $Z(t)$ by constructing the following candidate feasible point $(\mathcal{J}^{\mathcal{E}_1}, \mathcal{J}^{\mathcal{E}_2}, x, y)$, by setting

$$\begin{aligned}
\mathcal{J}^{\mathcal{E}_1} &= \lambda \mathcal{J}^{\overline{\mathcal{E}_1}} + (1-\lambda) \mathcal{J}^{\tilde{\mathcal{E}_1}} \\
\mathcal{J}^{\mathcal{E}_2} &= \lambda \mathcal{J}^{\overline{\mathcal{E}_2}} + (1-\lambda) \mathcal{J}^{\tilde{\mathcal{E}_2}} \\
x &= \lambda \overline{x} + (1-\lambda) \tilde{x} \\
y &= \lambda \overline{y} + (1-\lambda) \tilde{y}
\end{aligned} \tag{51}$$

We now verify this solution is feasible. It is clear that the newly constructed Choi operators are positive semidefinite. Furthermore we do have by construction that $x - y = 1$, and $\text{Tr}_{\overline{B}}[\mathcal{J}^{\mathcal{E}_1}] = xI_B$ and $\text{Tr}_{\overline{B}}[\mathcal{J}^{\mathcal{E}_2}] = yI_B$. Lastly, $\text{Tr}_{B_i}[\mathcal{J}^{\mathcal{E}}] = \lambda \Lambda^{\bar{t}} + (1-\lambda)\Lambda^{\tilde{t}} = \Lambda^t$.

Now, we show the first point. Suppose $t > 1$. Say $t = 1 + s$ where $s > 0$. Then we have that

$$\begin{aligned}
\|\Lambda_t - id\|_\diamond &= (1+s)\|\mathcal{R} - id\|_\diamond \\
&\geq \|\mathcal{R} - id\|_\diamond
\end{aligned} \tag{52}$$

Thus, if a point with $t = 1 + s$ is feasible, then $\|\Lambda_{(1+s)} - id\|_\diamond = (1+s)\|\mathcal{R} - id\|_\diamond \leq \delta$. Thus $\|\Lambda_{(1)} - id\|_\diamond = \|\mathcal{R} - id\|_\diamond \leq \delta$ So a point with $t = 1$ is feasible, which corresponds to just taking the fully depolarizing channel. Since this is a quantum channel, it a sample complexity of 1. Since the channel associated $t = 1 + s$ is not CP, its sample complexity is $> 1$. Thus we can without loss of optimality ignore points where $t > 1$. On the other hand, consider the case $t < 0$. Suppose $t = -s$ for some $s > 0$. Now

$$\begin{aligned}
\|\Lambda_t - id\|_\diamond &= s\|\mathcal{R} - id\|_\diamond \\
&\geq \|\mathcal{R} - id\|_\diamond
\end{aligned} \tag{53}$$

A similar arguement to above shows that point with $t < -1$ being feasible implies that $t = 1$ is feasible. Since the case where $t < -1$ is not CPTP, we can again without loss of optimality ignore points where $t < -1$.

Now we prove the second point, $Z(|t|) \leq Z(|t|)$ for $t \in [-1, 1]$. We write $|t| = \lambda(-|t|) + (1-\lambda)1$ where $\lambda = \frac{1-|t|}{1+|t|}$. Since $|t| \leq 1, \lambda \in [0, 1]$. So by convexity we have

$$\begin{aligned}
Z(|t|) &\leq \lambda Z(-|t|) + (1-\lambda)Z(1) \\
&\leq \lambda Z(-|t|) + (1-\lambda)Z(-|t|) \\
&= Z(-|t|)
\end{aligned} \tag{54}$$

Since $Z(-|t|) \geq Z(1) = 1$.

For the final point, suppose $0 \leq t_2 \leq t_1 \leq 1$. Then we can write $t_1 = \lambda t_2 + (1-\lambda)1$ for some $\lambda \in [0, 1]$. By convexity we have

$$\begin{aligned}
Z(t_1) &\leq \lambda Z(t_2) + (1-\lambda)Z(1) \\
&\leq \lambda Z(t_2) + (1-\lambda)Z(t_2) \\
&= Z(t_2)
\end{aligned} \tag{55}$$

where the first inequality follows from convexity and the second follows from $Z(t_2) \geq Z(1) = 1$. $\square$